\documentclass[%
 aip,
 amsmath,amssymb,
 reprint,%
]{revtex4-1}
\usepackage[english]{babel}
\usepackage{graphicx}
\usepackage{dcolumn}
\usepackage{bm}

\usepackage[utf8]{inputenc}
\usepackage[T1]{fontenc}
\usepackage{mathptmx}
\usepackage{etoolbox}
\usepackage{chemformula}
\usepackage{physics}

\makeatletter
\def\@email#1#2{%
 \endgroup
 \patchcmd{\titleblock@produce}
  {\frontmatter@RRAPformat}
  {\frontmatter@RRAPformat{\produce@RRAP{*#1\href{mailto:#2}{#2}}}\frontmatter@RRAPformat}
  {}{}
}%
\makeatother
\begin{document}

\preprint{AIP/123-QED}

\title{Eliminating Delocalization Error through Localized Orbital Scaling Correction with Orbital Relaxation from Linear Response}

\author{Yichen Fan}

\author{Jincheng Yu}%
\altaffiliation[Present address: ]{Department of Chemistry and Biochemistry, University of Maryland, College Park, MD 20742}
\author{Jiayi Du}

\author{Weitao Yang}
\altaffiliation{%
Department of Physics, Duke University, Durham, NC 27708
}%
\email{weitao.yang@duke.edu}
\affiliation{ 
Department of Chemistry, Duke University, Durham, NC 27708
}%

\date{\today}

\begin{abstract}
Despite the great success that Kohn-Sham density functional theory (KS-DFT) has achieved, the delocalization error remains a major challenge for commonly used density functional approximations (DFAs), resulting in systematic errors in ionization energies, electron affinities, band structures, and charge distributions.  A recently developed localized orbital scaling correction (LOSC) method, namely linear response LOSC (lrLOSC), addresses these challenges by incorporating a functional correction that includes the screening effect and orbital localization within the LOSC framework. The method has been shown to provide accurate descriptions of bulk systems and core-level binding energies in small molecular systems. In this work, we extend the applicability of lrLOSC to a broader range of molecular systems, spanning various sizes, with a focus on the corrections to valence orbital energies and total energies.  To enable the calculation of large chemical systems, we developed an efficient implementation of lrLOSC with computational costs comparable to standard KS-DFT calculations.  Numerical results show that, while screening provides modest improvements for small molecules, it becomes critical for achieving high accuracy in larger molecules, from linear to three-dimensional systems.  With the screening effect well captured in a unified way,  lrLOSC provides accurate descriptions for a wide range of chemical systems, including organic molecular systems of varying sizes and transition-metal oxide complexes, establishing it as a powerful tool for enhancing the reliability of computational simulations of chemical systems.
\end{abstract}

\maketitle

Kohn-Sham density functional theory (KS-DFT)\cite{hohenberg_inhomogeneous_1964, kohn_self-consistent_1965, parr_density-functional_1989} is highly regarded and widely
used to describe the electronic structures of molecules and bulk systems in chemistry,
physics, and materials science at a feasible computational cost. 
Driven by the need for accurate predictions from DFT-based theoretical simulations,
the development of density functional approximations (DFAs) 
has become a vibrant research area in quantum chemistry.
Over the past few decades,
conventional DFAs, including
local density approximations (LDAs)\cite{vosko_accurate_1980}, generalized gradient approximations (GGAs)\cite{becke_density-functional_1988, lee_development_1988, perdew_generalized_1996}, meta-GGAs\cite{beckeNewInhomogeneityParameter1998,taoClimbingDensityFunctional2003,peverati_m11-l_2012, sunAccurateFirstprinciplesStructures2016, yu_mn15-l_2016}, hybrid
functionals\cite{becke_density-functional_1988, lee_development_1988, stephens_ab_1994} and range-separated hybrid functionals \cite{savinDensityFunctionalsYukawa1995a,yanaiNewHybridExchange2004,cohenDevelopmentExchangecorrelationFunctionals2007a,mardirossian_omega_2016} have achieved significant success.

Despite their great success,
conventional DFAs suffer from major drawbacks: inaccurate valence orbital energies\cite{hybertsen_electron_1986, dabo_koopmans_2010, tsuneda_koopmans_2010, kronik_excitation_2012}, underestimated reaction barriers\cite{zheng_dbh2408_2009}, and inaccurate charge distribution descriptions\cite{li_localized_2018,mei_self-consistent_2020}, et al. These shortcomings originate from a systematic error known as the delocalization error\cite{mori-sanchez_localization_2008, cohen_insights_2008} (DE), which exhibits a size-dependent behavior. For small chemical systems with limited physical extent, 
DE manifests as a violation of the Perdew-Parr-Levy-Balduz (PPLB) condition\cite{
perdew_density-functional_1982, yang_degenerate_2000},
which states that the exact total energy $E(N)$, 
as a function of the number of electrons,
should be piecewise linear between consecutive integer values of $N$.
Energies from conventional DFAs are generally reliable at integer points, but display a convex deviation from the PPLB condition\cite{perdew_density-functional_1982, yang_degenerate_2000} at fractional electron numbers. 
As the size of the system increases,
this convex deviation decreases and disappears at the bulk limit. \cite{mori-sanchez_localization_2008} 
However, at the large system and bulk limit,
the DE manifests differently: 
conventional DFAs produce quantitatively incorrect total energies for charged systems at integer electron numbers\cite{mori-sanchez_localization_2008}.
To eliminate the systematic DE, many methods have been developed,
including 
range-separated hybrid functionals\cite{yanaiNewHybridExchange2004, santra_benefits_2022},
dielectric-dependent hybrid functionals\cite{brawand_generalization_2016},
self-interaction error corrected functionals\cite{perdew_self-interaction_1981, pederson_communication_2014},
DFT+U approaches\cite{anisimov_transition_2005, kulik_density_2006},
and Koopmans-compliant functionals\cite{nguyen_koopmans-compliant_2018, colonna_koopmans_2022}.

In addition to the aforementioned methods, a series of scaling correction (SC) methods\cite{zheng_improving_2011,  li_local_2015,zhang_orbital_2015, li_localized_2018, su_preserving_2020, mei_exact_2021, mei2022libsc, anPyGSCPythonTool2026} has been developed.
Extensive numerical studies have shown that SC methods effectively eliminate DE
and provide accurate numerical descriptions for numerous critical problems,
such as the photoemission spectroscopy\cite{li_localized_2018, mei_approximating_2018, yang_density_2020, mei_exact_2021},
photoexcitation energies\cite{mei_approximating_2018, meiChargeTransferExcitation2019, meiExcitedStatePotentialEnergy2019,li2022combining},
fundamental gaps\cite{zheng_improving_2011, li_localized_2018, mei_approximating_2018},
and polarizability\cite{li_local_2015, meiDescribingPolymerPolarizability2021}.

Building upon these successes, a novel SC method, namely linear-response localized orbital scaling correction (lrLOSC), has recently been developed. This method has demonstrated remarkable accuracy, as evidenced by its successful applications to bulk materials\cite{williams_correcting_2024} and core-level binding energies in small molecular systems\cite{yuAccuratePredictionCoreLevel2025}. The high accuracy of lrLOSC stems from its two critical components: orbital localization and orbital relaxation, the latter of which is effectively captured with the linear response theory.

In this work,
we extend lrLOSC’s applicability to molecular systems of all sizes,
highlighting its capability to provide accurate corrections to 
both total energies and valence orbital energies.
Specifically, we aim to evaluate the roles of orbital localization and orbital relaxation (or the screening effect) in molecular systems.
We begin by introducing the theoretical framework of lrLOSC, 
outlining its key components and implementation details.
We then present numerical results from our calculations, 
demonstrating the effectiveness and efficiency of lrLOSC in its current implementation.

In lrLOSC, the correction to the total energy, $\Delta E_{\text{lrLOSC}}$, 
defined as the difference between the total energy from lrLOSC, $E_{\text{lrLOSC}}$,
and the total energy from a DFA, $E_{\text{DFA}}$,
is given by 

\begin{equation}
    \label{eqn:totE_LOSC}
    \Delta E_\text{lrLOSC} \equiv E_{\text{lrLOSC}} - E_\text{DFA} = \frac{1}{2}\sum_\sigma \sum_{pq} \kappa_{pq}^\sigma \lambda_{pq}^\sigma ( \delta_{pq} - \lambda_{pq}^\sigma),  
\end{equation}
where $\kappa^\sigma$ is the curvature matrix for spin $\sigma$, 
and $\lambda^\sigma$ is the localized orbital (LO) occupation matrix,
$\lambda^\sigma_{pq}=\langle \phi_{p\sigma} | \hat{\rho} | \phi_{q\sigma} \rangle$.
In lrLOSC,
the LOs $\{\phi_{p\sigma}\}$ are also called orbitalets, 
and they are localized in both the physical space and the energy space.
Based on Eq.~(\ref{eqn:totE_LOSC}),
to obtain the energy correction from lrLOSC,
the two main tasks are
a) constructing the orbitalets and calculating the LO occupation matrix $\lambda^\sigma$
and
b) constructing the curvature matrix $\kappa^\sigma$.
We will describe how to complete the two tasks in lrLOSC respectively.

The localization procedure employed in lrLOSC is 
from the second version of LOSC (LOSC2)\cite{su_preserving_2020}.
Orbitalets are constructed through a unitary transformation of the canonical molecular orbitals (CMOs) $\ket{\psi^\sigma_n}$ in both the occupied and virtual subspaces for each spin channel $\sigma$ 
\begin{equation}
    \label{Unitary Transformation}
    \ket{\phi_p^\sigma} = \sum_n U^\sigma_{pn} \ket{\psi_n^\sigma},
\end{equation}
where the linear coefficients $U^\sigma_{pn}$ are determined by minimizing the cost function $F^\sigma$.
\begin{align}
\label{eqn:costFunc}
    F^\sigma &= (1-\gamma) \sum_{p} \left [\braket{\phi_{p}^{\sigma}|\boldsymbol{r}^2}{\phi_{p}^\sigma} - \braket{\phi_{p}^{\sigma}|\boldsymbol{r}}{\phi_{p}^\sigma}^2\right] \nonumber\\
    &+ \gamma C \sum_{p} \left [\bra{\phi_{p}^{\sigma}}\hat{h}^2\ket{\phi_{p}^\sigma} - \bra{\phi_{p}^{\sigma}} \hat{h}\ket{\phi_{p}^\sigma}^2 \right],
\end{align}
which represents a weighted sum of the physical-space cost function $\sum_{p}[\braket{\phi_{p}^{\sigma}|\boldsymbol{r}^{2}}{\phi_{p}^{\sigma}}-(\braket{\phi_{p}^{\sigma}|\boldsymbol{r}}{\phi_{p}^{\sigma}})^{2}]$ and the energy-space cost function $\sum_{p}[\bra{\phi_{p}^{\sigma}}\hat{h}^{2}\ket{\phi_{p}^{\sigma}}-(\bra{\phi_{p}^{\sigma}}\hat{h}\ket{\phi_{p}^{\sigma}})^{2}]$,
and $\hat{h}$ is the (generalized) KS one-electron Hamiltonian. 
In Eq. (\ref{eqn:costFunc}), $\gamma$ is the weight of the energy-space cost function, restricted to $0 \leq \gamma \leq 1$. This cost function includes contributions from both energy and spatial penalties; parameter C is introduced to convert the energy units to spatial units, and it is equal to 1000 in atomic units. Note that a similar cost function was developed by Gygi and others \cite{gygi_computation_2003, giustino_mixed_2006, dawes_using_2006}for bulk systems, but restricted to the occupied states (valence band) alone. In contrast, for both molecular and bulk systems, our localization procedure is applied to all states, including occupied and virtual, or equivalently the valence and conduction bands. The resulting LOs correspond to molecular orbitalets\cite{su_preserving_2020} in finite systems and to dually localized Wannier functions in periodic materials\cite{mahlerWannierFunctionsDually2025}.


The localization in Eq. (\ref{eqn:costFunc}) can be understood based on its limits. When $\gamma =1$, $F^{\sigma}$ encompasses solely the energy localization, making the orbitalets equivalent to CMOs.
When $\gamma =0$, all energy information is sacrificed to achieve the maximum localization 
in the physical space.
At this limit, $F^{\sigma}$ closely resembles the cost function of the Foster-Boys localization\cite{boys_construction_1960}; yet a major distinction arises: the linear combination in Eq. (\ref{Unitary Transformation}) extends to both occupied and virtual orbitals.
This inclusion allows the presence of fractionally occupied orbitals, further enabling  total energy corrections for physical systems with an integral number of electrons\cite{su_preserving_2020}.
In practical computations for lrLOSC, the parameter $\gamma=0.707$ is consistently applied across different systems \cite{li_localized_2018, su_preserving_2020}. 
This value leads to
optimal orbitalets with dynamic behavior,
 meaning that they adapt to the system's geometry and can be just canonical orbitals or space localized orbitals, depending on the one-electron energy spectrum at the specific geometry.

In small systems with limited physical extent, like a water molecule at its equilibrium geometry,
there is often a notable energy gap between the occupied orbitals and the virtual orbitals. This significant energy gap results in a high cost for the localization in the energy space when one attempts to mix occupied and virtual orbitals. Therefore, the derived orbitalets closely resemble canonical orbitals with very little mixing between occupied and virtual orbitals. On the other hand, as the system size increases or the chemical bonds are stretched, the energy gap between occupied and virtual orbitals decreases, enhancing the significance of spatial localization. This leads to spatially localized orbitals that can exhibit considerable fractional occupations within $\lambda_{pq}^{\sigma}$. Therefore, in smaller systems, the dynamic localization of orbitalets produces diagonal and integer values for $\lambda_{pp}^{\sigma}$, whereas for larger systems, fractional $\lambda_{pp}^{\sigma}$ values are commonly observed \cite{li_localized_2018, su_preserving_2020}.
In addition to serving as intermediate orbitals for computing lrLOSC corrections,
orbitalets can also qualitatively reveal the chemically reactive 
regions of various systems\cite{yu2022describing,long2024chemical}.

Now we introduce the second component of lrLOSC, namely the curvature matrix.
In lrLOSC, the curvature matrix $\kappa^\sigma$ employed in Eq.~(\ref{eqn:totE_LOSC})
is designed to include the screening effect or orbital relaxation.
In the initial version of LOSC\cite{li_localized_2018, su_preserving_2020}, the frozen orbital approximation was adopted as the global scaling correction (GSC)\cite{zheng_improving_2011}, providing accurate predictions for both small and large molecules. However, this approximation significantly overestimates the band gaps for bulk systems\cite{mahler_localized_2022}, making it unsuitable as a universal approach. Incorporating orbital relaxation has shown improvements in orbital energies\cite{zhang_orbital_2015, mei_exact_2021}. A clear and straightforward formulation of orbital relaxation in $\kappa_{pq}$ was initially presented for frontier orbitals during the earlier development\cite{yang_analytical_2012} of GSC, and was later extended to all canonical orbitals and to include cross terms by \citealt{mei_exact_2021}. The results of the Koopmans-compliant functional in orbital energies further underscores the importance of orbital relaxation. \cite{colonna_koopmans_2022} 

The expression of the curvature matrix in lrLOSC 
can be derived from several approaches,
including the coupled perturbed equations,
the combination of the Maxwell relation and linear response,
and using the density matrix as the direct variable,
as documented in the supplementary material (SM)\cite{SI} of Ref.~\citenum{mei_exact_2021}.
The form of the curvature matrix is
\begin{align}
    \label{eqn:kappa}
    \kappa_{pq}^\sigma= K_{pp\sigma, qq\sigma} - \sum_{ia\tau,jb\nu} (K_{pp\sigma, ia\tau} + K_{pp\sigma, ai\tau}) M^{-1}_{ia\tau, jb\nu}K_{jb\nu, qq\sigma},
\end{align} 
and
\begin{equation}
    \label{time-consuming}
    M_{ia\tau, jb\nu} = (\varepsilon_{a\tau} -\varepsilon_{i\tau}) \delta_{ij}\delta_{ab}\delta_{\tau\nu} +  (K_{ia\tau,jb\nu} + K_{ia\tau, bj\nu}),
\end{equation}
where $i, j$ are indices for occupied CMOs, $a, b$ are indices for virtual CMOs, $p,q$ are indices for orbitalets, and $\sigma, \tau, \nu$ are spin indices. The matrix $M_{ia\sigma, jb\tau}$ was derived by \citealt{yang_analytical_2012}.
In Eq.~(\ref{time-consuming}), the Hartree-exchange-correlation kernel matrix $K_{pq\tau,rs\nu}$ is
\begin{align}
    K_{pq\tau, rs\nu}
    &= \int d\textbf{r}_1d\textbf{r}_2d\textbf{r}_3d\textbf{r}_4\varphi_{p\tau}(\textbf{r}_1)\varphi^*_{q\tau}(\textbf{r}_2) \nonumber\\
    &\times K^{\sigma,\tau}(\textbf{r}_1, \textbf{r}_2;\textbf{r}_3, \textbf{r}_4) \varphi^*_{r\nu}(\textbf{r}_3)\varphi_{s\nu}(\textbf{r}_4),
\end{align}
where
\begin{align}
    K^{\sigma, \tau}&(\textbf{r}_1, \textbf{r}_2;\textbf{r}_3, \textbf{r}_4)  \nonumber \\
    &= \cfrac{\delta (\boldsymbol{r}_1 , \boldsymbol{r}_2) \delta (\boldsymbol{r}_3 , \boldsymbol{r}_4)}{|\boldsymbol{r}_1-\boldsymbol{r}_3|}+ \cfrac{\delta^2 E_{\text{xc}}}{\delta \rho^\sigma_s(\boldsymbol{r}_2 , \boldsymbol{r}_1) \delta \rho^\tau_s (\boldsymbol{r}_3 , \boldsymbol{r}_4)}.
\end{align}
$E_\text{xc}$ is the exchange-correlation (xc) energy evaluated from DFA.

With both the LO occupation matrix and the curvature matrix constructed,
we now have the expression for the orbital energies from lrLOSC
\begin{align}
    &\varepsilon_{i\sigma}^{\text{LOSC}}	
    =\frac{\partial E_{\text{LOSC}}}{\partial n_{i}^\sigma} \nonumber\\
	&=\varepsilon_{i\sigma}^{\text{DFA}}
    +\sum_{p}\left[\left(\frac{1}{2}-\lambda_{pp}^{\sigma}\right)\kappa_{pp}^{\sigma}\left|U_{pi}^{\sigma}\right|^{2}-\sum_{q\neq p}\kappa_{pq}^{\sigma}\lambda_{pq}^{\sigma}U_{pi}^{\sigma}U_{qi}^{\sigma*}\right],
\end{align}
where $U$ is the unitary matrix in Eq.~(\ref{Unitary Transformation}). Note that the $\gamma$ is the only parameter in the cost function, which can be used to directly correct the DE in any DFA without requiring additional tuning.


While the inclusion of screening effects significantly enhances the accuracy of lrLOSC, it also brings a considerable increase in computational cost compared to the original curvature matrix employed in
LOSC\cite{li_localized_2018, su_preserving_2020}. Specifically, constructing the curvature matrix $\kappa$ requires inverting the matrix $\mathbf{M}$ in Eq.~(\ref{time-consuming}), which demands extensive computational resources, both in terms of time and memory. The $\mathbf{M}$ matrix has a dimension of $N_{occ}N_{vir} \times N_{occ}N_{vir}$, where $N_{occ}$ designates the number of occupied orbitals, and $N_{vir}$ denotes the number of virtual states. Calculating the inverse of $\mathbf{M}$ directly makes lrLOSC an $\mathcal{O}(N^6)$ algorithm, with $N$ being the number of electrons. 
The high computational cost will restrict the application of lrLOSC to small and medium-sized systems.

To broaden the applicability of lrLOSC, 
in this study, we introduce an alternative way to compute $\kappa$,
which leverages the algebraic structure of the matrix $\mathbf{M}$ and utilizes a Resolution-of-Identity \cite{vahtrasIntegralApproximationsLCAOSCF1993,ren_resolution--identity_2012} approximation (RI-V).

The conventional RI-V scheme \cite{ren_resolution--identity_2012} decomposes the four-center electron repulsion integrals into products of two- and three-center integrals using the atomic-centered auxiliary basis functions $\{f_P(\textbf{r})\}$, such that:
\begin{equation}
    (ia|jb) = \sum_{PQ} C_{ia}^P V_{PQ}^{-1} C_{jb}^Q,
\end{equation}
where $V_{PQ}^{-1}$ is the inversion of the two-center Hartree integral
\begin{equation}
    V_{PQ} = \int d\mathbf{r} d\mathbf{r}' f_{P}(\mathbf{r})\frac{1}{|\mathbf{r} - \mathbf{r'}|} f_Q(\mathbf{r}),
\end{equation}
and $C_{ia}^P$ are the expansion coefficients which are chosen to be
\begin{equation}
    C_{ia}^Q = \sum_P (ia|P) V_{PQ}^{-1}.
\end{equation}

To accommodate the xc contribution in \textbf{K}, we extend the conventional RI-V scheme in lrLOSC as follows:
\begin{align}
\label{Eqn:genRI} 
    K_{ia\tau,jb\nu} &= \iint d\mathbf{r}d\mathbf{r'} \cfrac{\varphi_{i\tau}(\textbf{r})\varphi_{a\tau}^*(\textbf{r})\varphi_{j\nu}^*(\textbf{r}')\varphi_{b\nu}(\textbf{r}')}{|\textbf{r}-\textbf{r}'|} \nonumber\\
    &+ \varphi_{i\tau}(\textbf{r})\varphi_{a\tau}^*(\textbf{r}) f_\text{xc}^{\tau\nu} \varphi_{j\nu}^*(\textbf{r}')\varphi_{b\nu}(\textbf{r}')\\
    &\approx \sum_{PQRS} C_{ia\tau}^P V_{PQ}^{-1} (V_{QR} + f_{QR}^\text{xc}) V_{RS}^{-1} C_{jb\nu}^S,
\end{align} 
and $f_{\text{xc}}$ is the second functional derivative used in time dependent-DFT.

This RI-like approximation provides a compact, lower-rank representation for $\textbf{K}$ if the exchange-correlation functional is local or semi-local, allowing the matrix $\textbf{M}$ to be efficiently restructured as:
\begin{equation}
\label{decomposition}
    \mathbf{M} = \textbf{A} + \textbf{X}\cdot \textbf{Y}^T,
\end{equation}
where
\begin{align}
    A_{ia\tau,jb\nu} &= (\epsilon_{a\tau} - \epsilon_{i\tau}) \delta_{ij} \delta_{ab} \delta_{\tau\nu}, \\ 
    X_{ia\tau,Q} &= \sum_P C_{ia\tau}^P V_{PQ}^{-1}, \\ 
    Y_{Q,jb\nu} &=\sum_{RS} (V_{QR} + f_{QR}^\text{xc}) V_{RS}^{-1} C_{jb\nu}^S.
\end{align}

By leveraging the decomposition in Eq.~(\ref{decomposition}), we apply the  \textit{Sherman-Morrison-Woodbury} (SMW) formula \cite{william_h_press_sparse_1988}, which enables the efficient inversion of matrix $\textbf{M}$ as:
\begin{equation}
    \textbf{M}^{-1} = \textbf{A}^{-1} - \textbf{A}^{-1} \textbf{X} \left( \textbf{I} + \textbf{Y}^T \textbf{A}^{-1} \textbf{X} \right)^{-1} \textbf{Y}^T \textbf{A}^{-1}.
\end{equation}
Here, \textbf{A} is diagonal, and \textbf{X} and \textbf{Y} are lower-rank three-center matrices with dimensions $(N_\text{occ} N_\text{vir})\times N_\text{aux}$, where $N_\text{aux}$ represents the number of auxiliary functions in $\{f_P\}$, and $N_\text{aux}$ is significantly smaller than $N_\text{occ}N_\text{vir}$. 
With the low-rank approximation, the computational cost of lrLOSC scales as $\mathcal{O}(N_\text{aux}^3  N_\text{LO})$, where $N_\text{LO}$ is the number of localized orbitals constructed by the LOSC2 localization scheme. This expense can be further reduced by imposing an energy window that restricts the orbitals participating in the localization. Because the LOSC2 objective function includes an energetic penalty term, one can select a relatively narrow energy window centered on the frontier orbitals; this only slightly degrades the quality of the orbitalets while substantially decreasing $N_\text{LO}$ and hence the prefactor of the total cost, which effectively remains at an essentially $\mathcal{O}(N^3)$ scaling.
This setup provides an efficient compromise between computational cost and memory usage. As a result, evaluating the lrLOSC curvature becomes practical for larger molecular systems. Further details on this RI implementation and the associated time-scaling analysis can be found in the first section of the SM.\cite{SI} 

With the new implementation,
we test the capability of lrLOSC for providing accurate corrections to
the valence orbital energies
and the total energies of molecular systems of different sizes.
The frontier orbital energies from lrLOSC offer valuable insights into the chemical potential, $\left(\pdv{E}{N}\right)_v$, which describes the energy change associated with the electron addition and removal processes. Therefore, frontier orbital energies are directly connected to the ionization potential (IP) and electron affinity (EA). \cite{cohen_fractional_2008, yang_derivative_2012, mei_approximating_2018} 
In this study, we compare the valence orbital energies from lrLOSC with those obtained from the global scaling correction with the analytical and exact second-order corrections (denoted as GSC2)\cite{mei_exact_2021}, which incorporates the screening effect based on CMOs, and LOSC2 \cite{su_preserving_2020}, which employs orbitalets but captures no orbital relaxation.
Through these comparisons,
we show the importance of the two theoretical components of lrLOSC: orbital localization and the electron screening effect.
We also compare the results from lrLOSC with those from $\Delta$SCF, 
which evaluates the first IP and EA through the total energy difference between the neutral species and the ($N\pm1$)-electron systems.

To gain a comprehensive picture of the performance of lrLOSC, considering that DE exhibits 
different behaviors as the molecular size increases, lrLOSC was tested on molecules of
varying sizes \cite{shahi_second-order_2009, richard_accurate_2016, curtiss_assessment_1997} as well as in gas-phase transition metal monoxide molecules. The complete test set is divided into four categories: small molecules (with no more than 8 atoms), large non-linear molecules (containing more than eight atoms, excluding polymers), polyacetylene (\ch{H(HC=CH)_nH}, $n\in [1, 10]$), and the transition metal monoxides. Please refer to the SM\cite{SI} for more details on the systems.

\begin{figure}
    \centering
    \includegraphics[width = \linewidth]{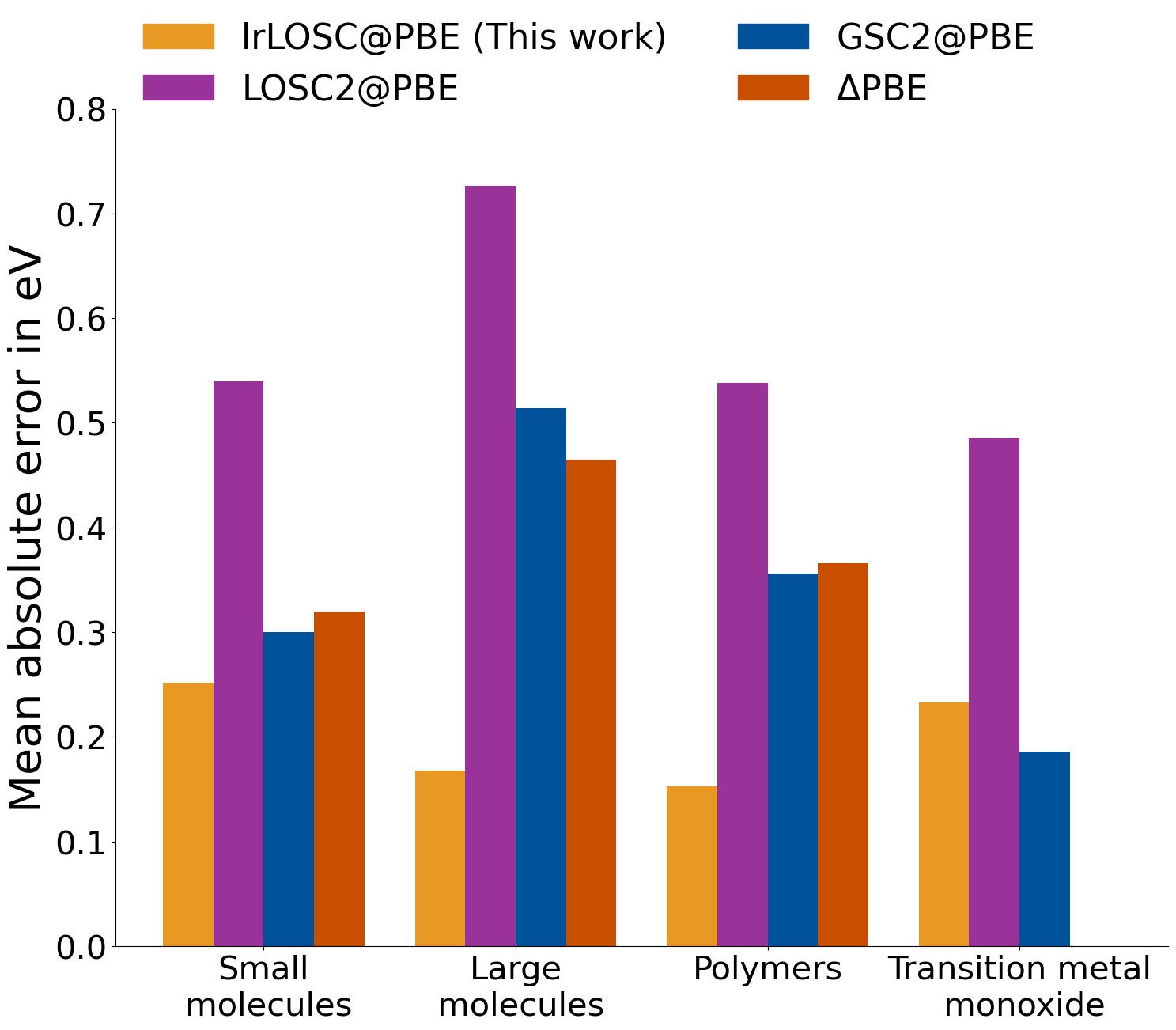}
    \caption{MAEs of IPs in eV, for gas-phase molecules of varying sizes obtained through different methods. The test set comprises 61 small molecules (containing no more than 8 atoms), 27 large non-linear molecules (with more than 8 atoms), poly-acetylene \ch{H(HC=CH)_nH} (where n ranges from 1 to 10), and transition metal monoxides from the fourth period. Note that, orbital energies from the parent GGA functional typically deviate by more than 4 eV.} 
    \label{fig:IE Bar plot}
\end{figure}

\begin{figure}
    \centering
    \includegraphics[width = \linewidth]{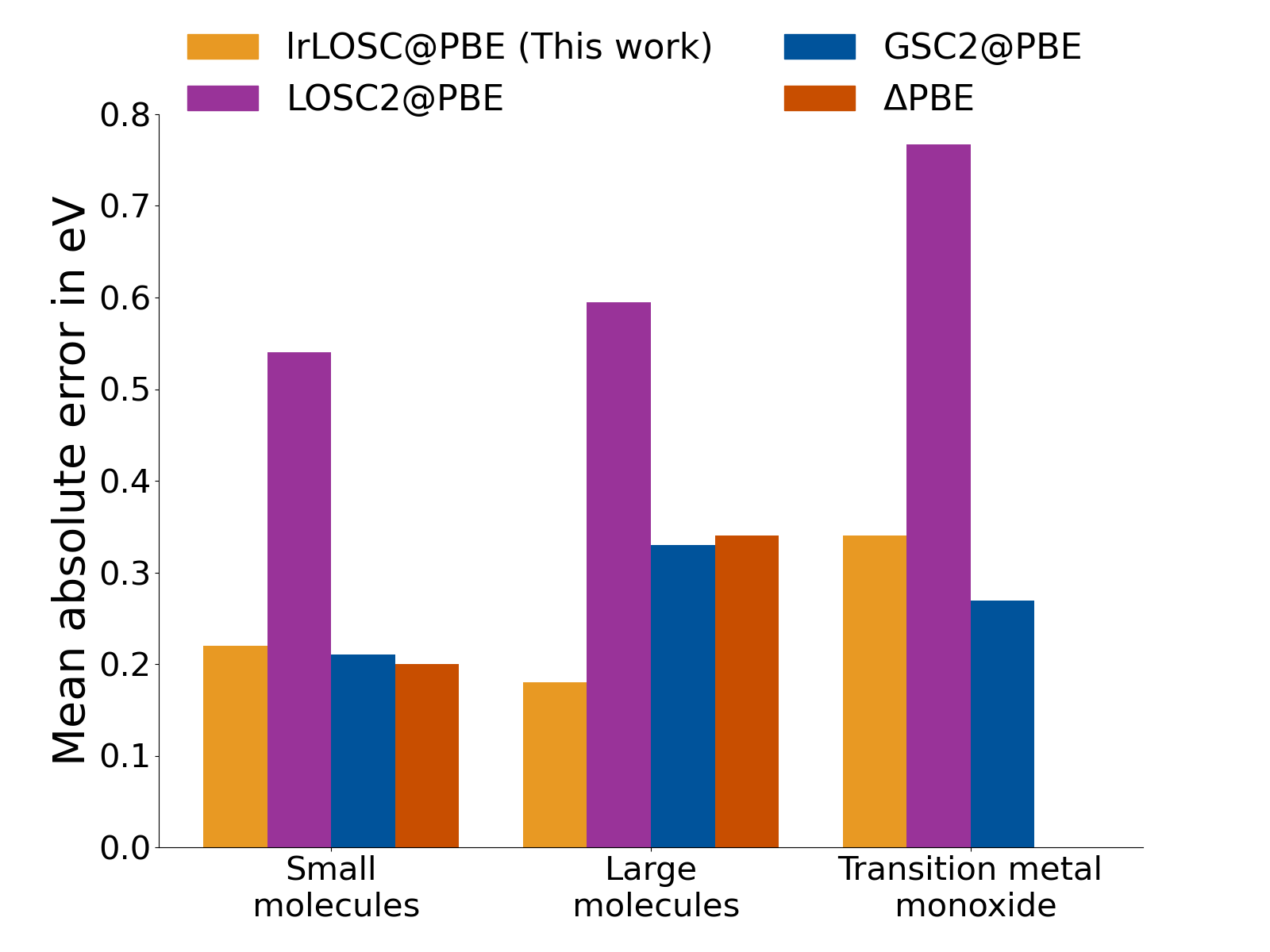}
    \caption{MAEs of EAs in eV, for gas-phase molecules of varying sizes obtained through different methods. The test set comprises 46 small molecules (containing no more than 8 atoms), 23 large non-linear molecules (with more than 8 atoms), and transition metal monoxides from the fourth period.  Note that, orbital energies from the parent GGA functional typically deviate by nearly 3.5 eV.}
    \label{fig:EA Bar plot}
\end{figure}

In Figure \ref{fig:IE Bar plot}, the mean absolute errors (MAEs) for IP are plotted across different methods. All calculations were performed based on the PBE functional \cite{perdew_generalized_1996}.
All lrLOSC, LOSC2, and GSC2 calculations are performed as a post-SCF
calculation. 
In Figure \ref{fig:EA Bar plot}, a similar comparison is shown for the EA. 
For the transition metal monoxides,
the ($N\pm1$)-electron system may be poorly defined, so the $\Delta$SCF results
are not provided. 
From  Figure~\ref{fig:IE Bar plot} and Figure~\ref{fig:EA Bar plot},
one can notice
that lrLOSC consistently shows the lowest MAEs among the four methods studied and that the errors
from lrLOSC
remain stable as the system size increases. 
The high accuracy of lrLOSC for systems of all sizes is due to
a) utilizing properly localized orbitalets
and 
b) the well described screening effect.
The role of orbital localization can be understood by comparing lrLOSC with GSC2.
While the two methods achieve similar accuracy for small molecules,
lrLOSC demonstrates significant 
improvement over GSC2 in the computation of large molecules and polymers, 
where a significant deviation between LO and CO is anticipated in these larger systems. 
Furthermore, 
to evaluate the significance of incorporating the screening effect, we compare the results of lrLOSC with those of LOSC2. 
While both methods utilize an identical localization procedure, 
lrLOSC further accounts for orbital relaxation through linear response theory.
With the screening effect well captured,
lrLOSC consistently achieves lower MAEs than LOSC2 
across molecules of all sizes.

In this work, we also apply our lrLOSC method for the first time to transition metal systems, which constitute a broad class of strongly correlated materials. lrLOSC yields accurate IP and EA predictions for gas-phase transition metal monoxides and reduces the error of the original LOSC approach by about half, highlighting the crucial role of the screening effects incorporated through linear response theory in transition metal calculations. Nonetheless, because the systems considered here are diatomic molecules, orbital localization effects are minor, and the orbital energy predictions obtained from lrLOSC and GSC2 are therefore very similar.

\begin{figure}

        \centering
        \includegraphics[width=1\linewidth]{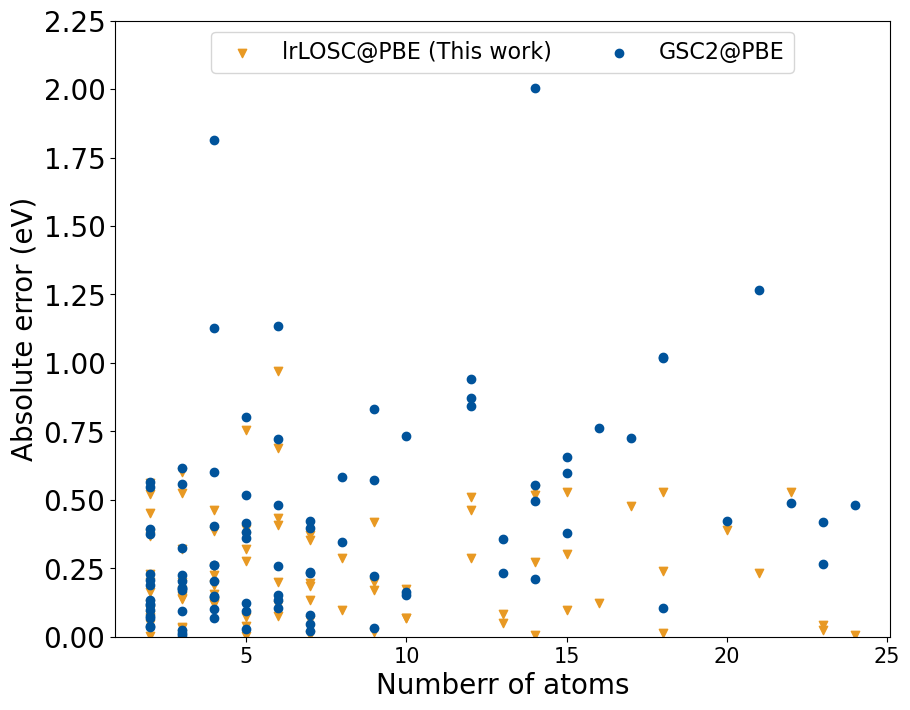}
        \caption{Error distribution of the first IP obtained from lrLOSC@PBE and GSC2@PBE for molecules with different sizes.}
    \label{fig:IP error plot}
\end{figure}

Figure \ref{fig:IP error plot} illustrates the absolute deviation in IP from the
reference values. For small systems, the discrepancy between the MAEs from lrLOSC and GSC2 is
negligible. However, for systems containing more
than 8 atoms, lrLOSC demonstrates superior performance over the GSC2 method.

lrLOSC shows promise in computing valence orbital energies. However, practical applications have
exposed certain limitations of lrLOSC. Previous research indicates that in most conventional DFAs, it is primarily the quadratic component that leads to deviations from the piece-wise linearity condition\cite{zheng_improving_2011,hait_delocalization_2018}. Nevertheless, contributions from higher-order expansion can have a significant impact under some scenarios. An evidential example of these contributions from the higher-order expansion is the emergence of a negative-curvature region when a small fraction of
an electron is added to certain integer systems. \cite{li_piecewise_2017} This phenomenon is rare, but it does occur.
The analytical expression of kappa
in Eq.~(\ref{eqn:kappa}) can capture these unstable negative curvatures and lead to an unphysical
energy correction. 

To mitigate issues associated with negative curvature and enhance the robustness of lrLOSC for a wide range of systems, we developed an adjusted electron density to sidestep instability in the second derivative. For molecular orbitals exhibiting problematic elements in the curvature matrix, we introduced a slight increment in electron density when assessing the functional derivative of the xc functional. This addition assists the functional derivative in bypassing its unstable region while having minimal impact on other orbitals. For the sake of consistency and clarity, we apply this strategy to all molecules and to each virtual orbital. Further details on convergence tests and implementations are available in the SM\cite{SI}.

\begin{figure}
        \centering
        \includegraphics[width=1\linewidth]{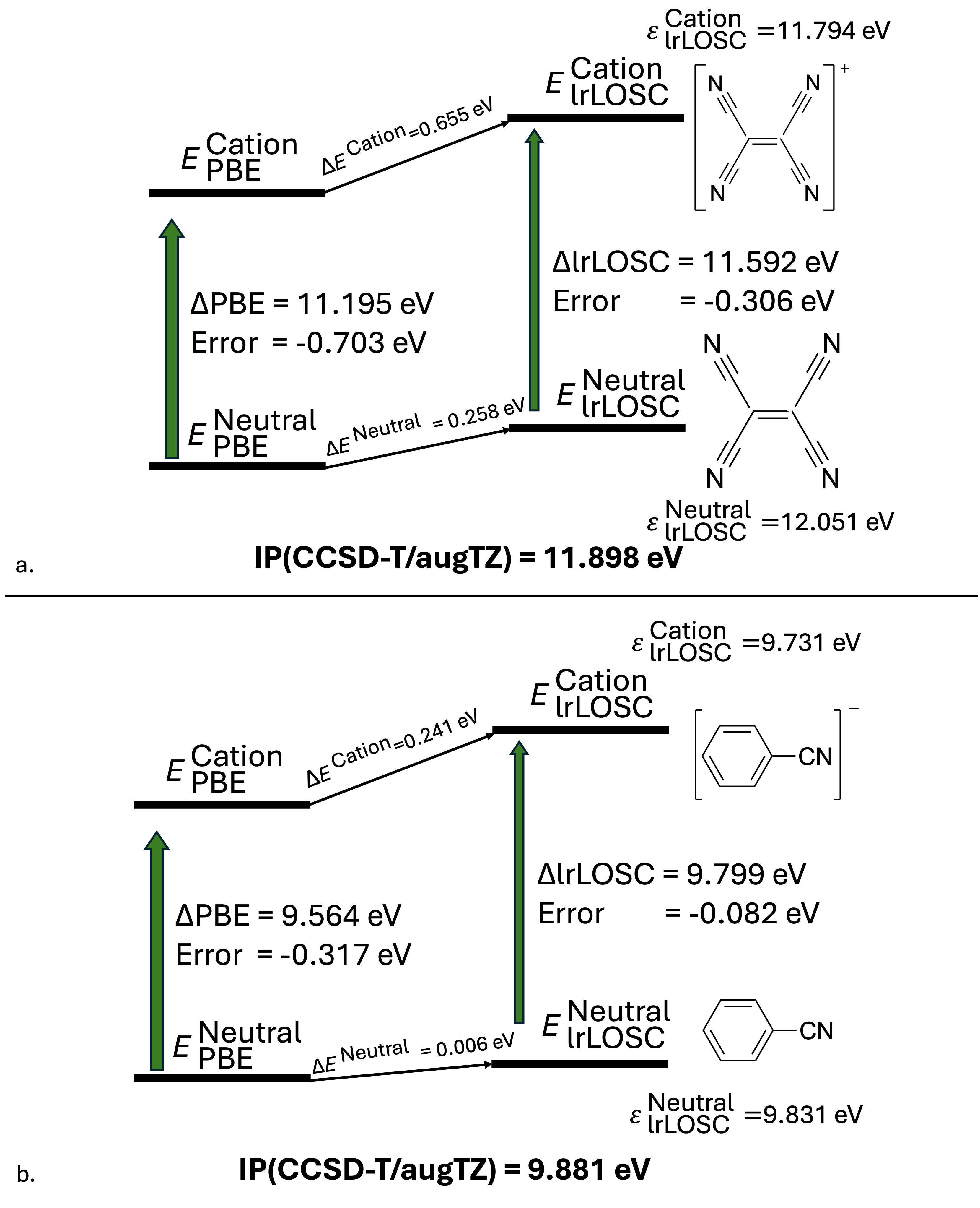}
\caption{lrLOSC total energy correction (\(\Delta E_{\text{lrLOSC}}\)) for (a) TCNE and its cation, and (b) benzonitrile and its cation. The vertical IPs are evaluated using three approaches:  
1. The total energy difference between the cationic and neutral systems (indicated by green arrows),  
2. The HOMO energy of the neutral system (bottom), and  
3. The LUMO energy of the cationic system (top).  
The reference IP values are obtained from EOM-CCSD(T) calculations using the aug-cc-pVTZ basis set \cite{dunning_gaussian_1989}. Errors relative to the EOM-CCSD(T) IP are reported beneath the total energy difference.}
    \label{fig:DlrLOSC}
\end{figure}

In addition to the valence orbital energies,
accurately assessing the total energies is also crucial, particularly for larger systems where DE undermines the reliability of DFAs.
Here, we demonstrate the effectiveness of lrLOSC in providing accurate total and orbital energies.
Figure \ref{fig:DlrLOSC} shows the vertical IPs of tetracyanoethylene (TCNE) and benzonitrile calculated using three methods, compared with EOM-CCSD(T) benchmarks. lrLOSC provides positive energy corrections, particularly for ($N-1$)-electron cationic systems, aligning the IPs more closely with EOM-CCSD(T) and highlighting the prominence of DE in charged systems. Table \ref{tab:total_Energy} reinforces this trend, showing systematic improvements in vertical IPs for 21 molecules, with a 0.19 eV reduction in MAE for lrLOSC-corrected $\Delta$SCF compared to uncorrected methods. These results demonstrate lrLOSC as a robust method for addressing DE in both orbital and total energy calculations across diverse molecular systems.

\begin{table}[]
    \centering
    \begin{tabular}{c|ccc}
    \hline
         & $\varepsilon_\text{HOMO}^\text{Neutral}$ & $\Delta$PBE & $\Delta$lrLOSC@PBE \\
         \hline
     MAE    & 0.19  & 0.51  & 0.32 \\
     MSE    & -0.17 & -0.51  & -0.32\\
     \hline
     
    \end{tabular}\\
    \vspace{0.2cm}
    \begin{tabular}{c|cc}
    \hline
         & $\Delta E^\text{Cation}$ & $\Delta E^\text{Neutral}$ \\
         \hline
        Mean absolute correction & 0.21 & 0.03\\
        Mean signed correction & 0.21 & 0.03\\
        \hline
    \end{tabular}
    \caption{Comparison of vertical IPs for 21 large molecules derived from three methods. Both MAE and MSE are relative to EOM-CCSD(T)/aTZ computational results\cite{richard_accurate_2016}. In the bottom table,  the average lrLOSC total energy corrections to the neutral molecules and cations are reported in the last two columns labeled with $\Delta E^\text{Neutral}$ and $\Delta E^\text{Cation}$,  which implies lrLOSC has more significant total energy correction to the open-shell cationic systems than the closed-shell neutral molecules. }
    \label{tab:total_Energy}
\end{table}

In summary, we present a universal treatment for the delocalization error named lrLOSC that provides estimates
of the valence orbital spectrum with reliable accuracy across molecular
systems of different sizes at a low computational cost. 
The lrLOSC method addresses the DE by focusing on
two principal aspects: orbital localization and the screening effect.
Numerical findings suggest that lrLOSC offers accurate estimations of
valence orbital energies and total energies for molecular systems 
of various sizes.
The critical roles of orbital localization and describing the screening effect
are understood by comparing results from lrLOSC and those from GSC2
and LOSC2.
The localization procedure is essential for
extended molecular systems,
and describing the screening effect improves the descriptions of
all molecular systems. Additionally, in this paper, we
explore the algebraic structure of lrLOSC, significantly enhancing its
efficiency and achieving a computational cost comparable to that of the
standard KS-DFT. 

\section*{Conflict of Interest}

The authors have no conflicts to disclose.

\section*{Supplementary Material}

Details of the Sherman–Morrison–Woodbury based lower rank approximation, validation of the small fraction treatment, computational details, and the complete benchmark data for all systems studied are included in the supplementary material\cite{SI} (See also references \cite{william_h_press_sparse_1988,QM4D,mokit,hattig_optimization_nodate, gilbert_self-consistent_2008,pasinszki_ground_2010, weigend_efficient_2002,clark_photoelectron_1973, eland_photoelectron_1969,meot-ner_ion_1980,liIntroducingGPUaccelerationPythonbased2024,wuEnhancingGPUaccelerationPythonbased2024}) therein.

\begin{acknowledgments}
JY and WY acknowledge support from the National Institutes of Health (1R35GM158181).
\end{acknowledgments}

\section*{Author Contributions}

Y.F. and J.Y. contributed equally to this paper.

\section*{Data Availability Statement}

The data that support the findings of this study are available within the article and its supplementary material.

\bibliography{lrLOSC_zotero,extra}

\end{document}